\documentclass[iop,apjl]{emulateapj}

\usepackage{apjfonts,graphicx}
\usepackage{wasysym,gensymb,amstext,epstopdf, amssymb}
\usepackage{textcomp,epsf,url}

\newcommand{\rproj}{$R_{proj}$}

\shorttitle{M31 Halo GC Kinematics}
\shortauthors{Veljanoski et al.}

\begin{document}

\title{Kinematics of Outer Halo Globular Clusters in M31}

\author{J. Veljanoski\altaffilmark{1}, 
A.M.N. Ferguson\altaffilmark{1,11}, 
A.D. Mackey\altaffilmark{2}, 
A.P. Huxor\altaffilmark{3}, 
M.J. Irwin\altaffilmark{4}, 
P. C\^{o}t\'{e}\altaffilmark{5,11}, 
N.R. Tanvir\altaffilmark{6},
E. J. Bernard\altaffilmark{1}, 
S. C. Chapman\altaffilmark{4}, 
R. A. Ibata\altaffilmark{7}, 
M. Fardal\altaffilmark{8}, 
G. F. Lewis\altaffilmark{9}, 
N. F. Martin\altaffilmark{7,10}, 
A. McConnachie\altaffilmark{6} 
and J. Pe\~{n}arrubia\altaffilmark{1}}

\altaffiltext{1}{Institute for Astronomy, University of Edinburgh, Royal Observatory, Blackford Hill, Edinburgh, EH9 3HJ, UK}
\altaffiltext{2}{Research School of Astronomy \& Astrophysics,Australian National University, Mt Stromlo Observatory, Weston Creek, ACT 2611, Australia} 
\altaffiltext{3}{Astronomisches Rechen-Institut, Zentrum fur Astronomie der Universitat Heidelberg, Monchhofstr. 12-14,69120 Heidelberg, Germany} 
\altaffiltext{4}{Institute of Astronomy, University of Cambridge, Madingley Road, Cambridge, CB3 0HA, UK} 
\altaffiltext{5}{NRC Herzberg Institute of Astrophysics, 5071 West Saanich Road, Victoria, British Columbia V9E 2E7, Canada}
\altaffiltext{6}{Department of Physics \& Astronomy, University of Leicester, Leicester, LE1 7RH, UK} 
\altaffiltext{7}{Observatoire de Strasbourg, 11, rue de l'Universit\'e, F-67000 Strasbourg, France}
\altaffiltext{8}{University of Massachusetts, Department of Astronomy, LGRT 619-E, 710 N. Pleasant Street, Amherst, Massachusetts, 01003-9305, USA }
\altaffiltext{9}{Institute of Astronomy, School of Physics, University of Sydney, NSW 2006, Australia}
\altaffiltext{10}{Max-Planck-Institut fuer Astronomie, Koenigstuhl 17, D-69117 Heidelberg, Germany}
\altaffiltext{11}{Visiting Astronomer, Kitt Peak National Observatory, National Optical Astronomy Observatory, 
which is operated by the Association of Universities for Research in Astronomy (AURA) under cooperative agreement with the National Science Foundation.}

\begin{abstract}
  We present the first kinematic analysis of the far outer halo
  globular cluster (GC) population in the Local Group galaxy M31. 
  Our sample contains 53 objects with projected radii of $\sim 20-130$~kpc, 
  of which 44 have no previous spectroscopic information. GCs with projected 
  radii $\gtrsim~30$ kpc are found to exhibit net rotation around the minor
  axis of M31, in the same sense as the inner GCs, albeit with
  a smaller amplitude of 79\textpm19 km/s. The rotation-corrected
  velocity dispersion of the full halo GC sample is 106\textpm12 km/s, 
  which we observe to decrease with increasing projected radius. 
  We find compelling evidence for kinematic-coherence amongst GCs which project 
  on top of halo substructure, including a clear signature of infall for GCs lying along 
  the North-West stream. Using the tracer mass estimator, we estimate the dynamical 
  mass of M31 within 200~kpc to be $M_{\rm M31} = (1.2-1.5) \pm 0.2 \times 10^{12}M_{\odot}$.
  This value is highly dependent on the chosen model and assumptions within.
  \end{abstract}


\section{Introduction}

Globular cluster (GC) systems contain important clues about the
assembly history of galaxies \citep[e.g.,][]{West04Nature}. Their 
kinematics are especially important as different formation channels
lead to distinct predictions \citep[e.g.][]{Forbes97,BrodieStraderReview}.
Moreover, GC kinematics can also be used to model the shape of the
gravitational potential and constrain the total mass of the host
galaxy \citep[e.g,][]{Cote01}.

Located at a distance of $\sim$780~kpc \citep{McConnachie05}, the Local 
Group galaxy M31 provides an excellent opportunity to study a rich GC 
system in unparalleled detail. It has more than 500 confirmed members 
listed in the Revised Bologna Catalogue \citep[RBC,][]{Galleti04RBC}, 
most of which lie within a projected radius (\rproj) of 30~kpc. In recent 
years, state-of-the-art wide field surveys \citep[e.g,][]{Ferguson02,Ibata07,McConnachie09} 
have enabled searches for GCs in M31's far outer halo. This has led to 
the discovery of over 90 new halo GCs, extending to \rproj\ $\sim140$~kpc and 3D radii 
of $\gtrsim$200~kpc (e.g, Huxor et al. 2005; Huxor et al. 2008; Huxor et al 2013, 
di Tullio Zinn \& Zinn 2013). A major step forward in understanding the formation 
of the outer halo GC system came from the realisation that these objects 
preferentially lie on stellar streams and other debris features 
\citep{Mackey10ApJL,Mackey13}. Monte Carlo simulations indicate a probability
of $\lesssim 1\%$ that such alignments should happen by chance,
leading to the conclusion that 80\% of the M31 outer halo GCs have
been accreted along with their host galaxies, confirming the idea 
put forward by \citet{SZ78} for the Milky Way. In this Letter, we present 
the first results of a spectroscopic survey of these outer halo objects, 
focusing on their global kinematics. A detailed description of the data 
as well as a full analysis is deferred for a later publication 
(Veljanoski et al. in prep).

\section{The data}

Spectra were acquired for
53 GCs spanning \rproj\ $\sim20-130$~kpc, of which 12(6)
lie beyond $\gtrsim80(100)$~kpc. This sample is complete 
down to $g\sim18.5$, and 44 of the clusters 
had not previously been observed spectroscopically.  The data were 
obtained over 15 nights during 2005-2010 using the ISIS 
spectrograph on the WHT 4.2m, and the RC spectrograph 
on the KPNO 4m. ISIS has two detectors that independently 
sample the blue and red spectral range. We used the R600B 
and R600R gratings to cover the wavelength range
$\sim350-510$~nm with a dispersion of 0.045~nm/pixel and
$\sim750-920$~nm with a dispersion of 0.079~nm/pixel
respectively. With the RC spectrograph, we used the KPC007 grating
with a wavelength coverage of $\sim350-650$~nm and a dispersion
of 0.139~nm/pixel. Total integrations were 600-7200 seconds
depending on the target brightness. The slit width was 1-2$''$. 
The signal-to-noise per pixel was $\approx 7-30$ for most 
targets and 50-70 for the brightest objects.

The data were reduced using standard IRAF\footnote{IRAF is distributed
  by the National Optical Astronomy Observatories, which are operated
  by the Association of Universities for Research in Astronomy, Inc.,
  under cooperative agreement with the National Science Foundation}
procedures. One-dimensional spectra were extracted with aperture radii 
of 2-2.5$''$. Heliocentric radial velocities were derived using a 
chi-squared minimization technique between GC spectra and radial 
velocity template stars (Veljanoski et al. 2013, in prep). This is 
analogous to the standard cross-correlation method, and it produces 
similar results. The method has the advantage that it uses the 
uncertainties in both the target and template spectra, which helps 
to eliminate spurious features in the chi-squared function. 

The uncertainty in the radial velocity of each cluster is adopted 
to be the standard deviation of all the independent chi-squared 
minimizations between the cluster and multiple radial velocity
standard stars. Furthermore, as we obtained two independent 
velocity measurements for the GCs observed with ISIS, these 
were combined to reduce the uncertainty in the final velocity value. 
The final median uncertainty of all 53 GC velocity measurements 
is 12~km/s. Of the 9 GCs in our sample that had published velocities 
in the RBC, all agree to within one standard deviation. Four of these 
clusters were found to have more precise RBC velocities compared 
to our measurements and for these we adopt the RBC values in our 
subsequent analysis.

As our GC sample spans a large extent on the sky, we converted our
heliocentric radial velocities to the Galactocentric frame in order to remove any
effects the Solar motion could have on the kinematics. This conversion
was done using the relations presented in \citet{Courteau99}, with
updated values for the Solar motion from \citet{McMillan11} and
\citet{Schonrich10}. For the purpose of this study, we take the M31
heliocentric velocity to be -301 \textpm 4~km/s \citep{Courteau99},
which translates to a Galactocentric radial velocity of
-109 \textpm 4~km/s.

\section{Results}

\subsection{The Global Velocity Map}

\begin{figure*}[t]
\begin{center}
\end{center}
\includegraphics[width=168mm]{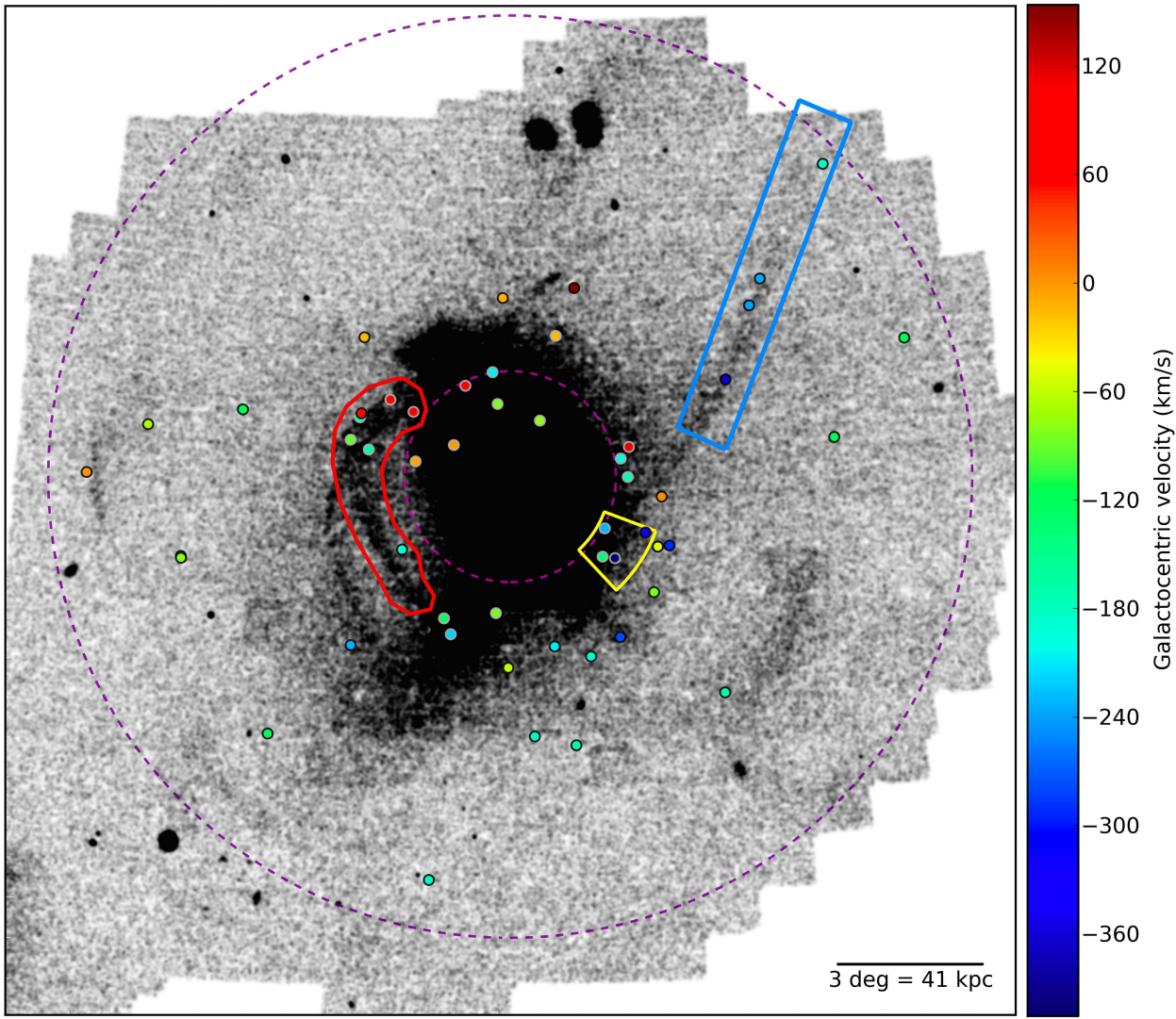}
\caption{The metal-poor stellar density map of M31 from PAndAS.
  Positions of the observed GCs are marked with colored dots 
  which correspond to their Galactocentric radial velocities in 
  units of km/s. Some GC subgroups are indicated (see text for
  details). The purple dashed circles correspond to 30 and 130 
  kpc radii in projection. The Galactocentric velocity of M31 is 
  -109 \textpm 4~km/s. North is up and East is left.}
\label{fig:velmap}
\end{figure*}

Figure \ref{fig:velmap} shows the most recent metal-poor ([Fe/H] $\lesssim -1.4$)
red giant branch stellar density map from the Pan-Andromeda Archaeological Survey
(PAndAS) \citep{McConnachie09}. Overlaid are the positions of 
the observed GCs, the colors of which correspond to their measured 
Galactocentric radial velocities. 

Three interesting groups of GCs are indicated on Figure
\ref{fig:velmap}. The blue rectangle marks 4 GCs that project on the
North-West stream. A contiguous velocity gradient is seen along this
feature, with the most radially-distant cluster (\rproj\ $\approx$125~kpc) 
having a velocity of $\approx-183$~km/s and the innermost object 
(\rproj\ $\approx$67~kpc) having $\approx-380$~km/s. Also
marked (red contour) are the GCs that lie on stream D, a feature first
identified by \citet{Ibata07}. There is no apparent velocity gradient
in this case, but the velocities of the GCs observed in the 
northeastern part of the stream suggest two distinct kinematic groups.
In particular, 4 GCs have a mean velocity and dispersion of
$-171$~km/s and 50~km/s, while the remaining 3 GCs have mean velocity
and dispersion of 71~km/s and 18~km/s. Interestingly, the northeastern
part of stream D has been previously identified to be a complicated
region where it overlaps stream C \citep{Ibata07, Richardson11}.
Thus, it seems likely that the two kinematic GC groups we have identified are
associated with these different streams, and are moving in the opposite sense 
around M31, as judged by their mean velocities. Radial velocities for stream
stars in this overlapping region are not yet available, but the lone
southern GC on stream D has a radial velocity that lies within
$\sim20$~km/s of stream stars tentatively identified in a nearby 
field \citep{Chapman08}. Interestingly, the And-I dwarf, located on 
the south end of Stream D, has a Galactocentric velocity of $\sim$-190~km/s, 
similar to one of the kinematic GC groups. The And-IX dwarf, located on the 
north end of Stream D with velocity of $\sim$-20km/s, appears uncorrelated with 
the GC groups.

The yellow region on Figure \ref{fig:velmap} marks
``association 2'' which \citet{Mackey10ApJL} identified as a
statistical overdensity of GCs not associated with any obvious
underlying stellar debris feature. The spread of velocities indicates 
that not all of the GCs in this region can be members of a kinematically 
coherent subgroup. However, measurements for the remaining 
$\sim$60\% of the putative ``association 2'' GCs are required before the presence 
of a subgroup can be definitely ruled out.

\subsection{Rotation and Velocity Dispersion}

\begin{figure}
\begin{center}
\end{center}
\includegraphics[width=86mm]{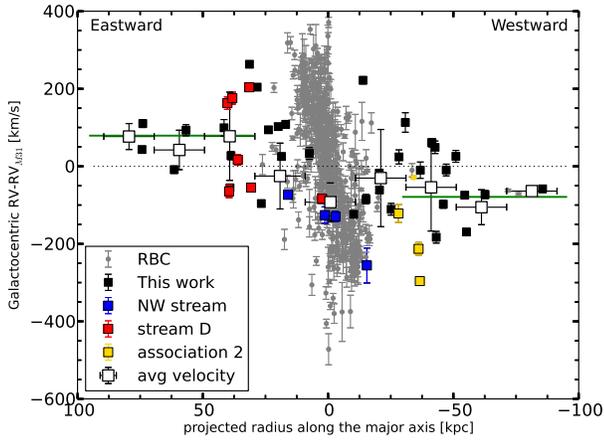}
\caption{Galactocentric radial velocity, corrected for the M31
  systemic motion, vs. projected radius along the M31 major axis. The
  black and colored squares mark the velocities of the GCs presented
  in this work, with different colors marking GC subgroups
  identified on Figure \ref{fig:velmap}. RBC values are shown as small
  grey circles. The open squares correspond to the mean velocities of 
  the GCs with \rproj \textgreater 30 kpc calculated in 20~kpc
  bins. The $x$ error bars represent the bin size, while the $y$ 
  error bars indicate the standard deviation of the mean. The green solid 
  lines correspond to our measured amplitude for the outer GCs corrected 
  for inclination.}
\label{fig:rot}
\end{figure}

It has been known for some time that GCs in the inner regions of M31
rotate around the minor optical axis of the galaxy, in the same sense
as the disk rotation. \citet{Perrett02} measured a rotational
amplitude of the GC system of $\sim140$~km/s, while \citet{Lee08} 
measured $\sim190$~km/s using a larger sample. Dividing the GCs based 
on their metalliticy, \citet{Deason11M31rot} found more pronounced rotation 
for the metal-rich ([Fe/H] $>$-1) subpopulation. Inspection of Figure
\ref{fig:velmap} strongly suggests this rotation persists to larger
radii, with GCs in the northeast having systematically higher
velocities than those in the southwest. Figure \ref{fig:rot} 
shows the Galactocentric radial velocities, corrected for the systemic 
motion of M31, versus their projected distances along the M31 major 
axis. GCs belonging to the subgroups identified in Figure \ref{fig:velmap} 
are color-coded. The rotational signature appears to be a property of the 
bulk population of the outer halo GC sample and is not driven by one or 
two kinematically-coherent subgroups.

To further investigate the rotational signature, we follow \citet{Cote01} 
and fit the observed projected Galactocentric radial velocities $v_p$ of 
the GCs with the function,

\begin{equation}
v_p(\theta) = v_{sys} + Asin(\theta - \theta_{0})
\label{eg:rot}
\end{equation}
where $\theta$ is the projected position angle, measured east of
north, of a cluster relative to M31 center, $\theta_{0}$ is the
projected position angle of the GC system rotation axis, $v_{sys}$ is
the systemic velocity of the GC system and $A$ is the amplitude of
rotation. This approach assumes that the rotation axis of the GC
system is perpendicular to the line of sight, and that the intrinsic angular
velocity of the system is constant on spherical surfaces. The
uncertainties are determined using the numerical bootstrapping
technique \citep{Efron82} and the derived rotational amplitudes are
corrected for the inclination of the M31 disk, taken to be 77.5\degree 
\citep{Ferguson02}.

We augment our radial velocities with those from the RBC and fit the
GC sample as a whole as well as within and beyond 30 kpc. 
This radius corresponds to a clear break
in the GC radial number density profile \citep{Huxor11} and therefore
provides a natural division between the ``inner'' and ``outer" halo.
For reference, the outer halo sample consists of the 48 GCs presented here, to
which we add a further 2 confirmed GCs from the RBC.

\begin{table}
 \caption{Derived Rotational Properties for M31 Halo GCs}
 \label{tab:1}
 \begin{tabular}{lllll}
 	\hline
	\hline
					& A			& $\theta_{0}$		& Velocity Dispersion	& $N_{GC}$	\\
					& [km/s]		& [degrees]		& [km/s]	 			\\
	\hline
	All GCs				& 133 \textpm 11 	& 124 \textpm 4 	& 115 \textpm 5		& 595		\\
	\rproj \textless 30 kpc		& 137 \textpm 10	& 124 \textpm 4		& 114 \textpm 5		& 545		\\
	\rproj \textgreater 30 kpc	& 79  \textpm 19 	& 123 \textpm 27	& 106 \textpm 12	& 50		\\
	\hline
 \end{tabular}
\end{table}

The systemic velocity of the GC system was set to the M31 Galactocentric 
systemic velocity. The results of this fitting are displayed in Table 
\ref{tab:1}. The results remain unchanged when $v_{sys}$ in Equation 
\ref{eq:rot} is left to vary as a free parameter, or when the mean 
velocity of the GC system is used. Given the position angle of the M31 
major axis is 38\degree, the derived rotation axis is consistent with 
the minor axis of M31, and is essentially indistinguishable from the 
rotation axis of the inner halo GCs. 

\begin{figure}
\begin{center}
\end{center}
\includegraphics[width=86mm]{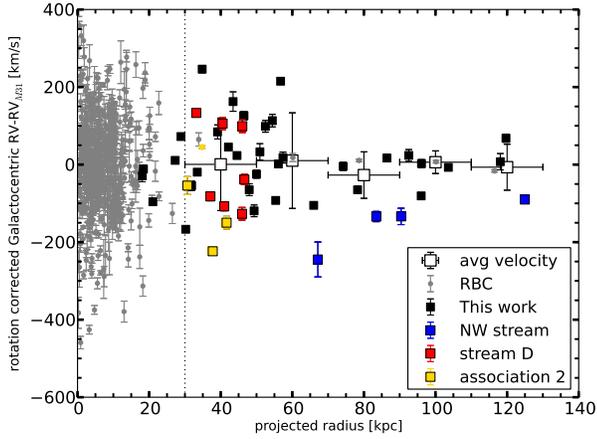}
\caption{Rotation-corrected Galactocentric radial velocity, corrected
for the M31 systemic motion, vs. \rproj\ from the M31 center. Symbols
  are as in Figure \ref{fig:velmap}. The open squares correspond to
  the mean velocities of the halo clusters in 20~kpc bins. The
  $x$ error bars mark the bin size, while the $y$ error bars represent 
  the rotation-corrected velocity dispersion. The dotted line marks the 
  30 kpc radius.}
\label{fig:vproj}
\end{figure}

Finally, we derived the rotation-corrected velocity dispersion of the
M31 GC sample using the biweight scale of \citet{Beers90}. Figure
\ref{fig:vproj} shows the rotation-corrected Galactocentric radial velocities 
(corrected for the systemic motion of M31) versus \rproj\ of the M31 GCs. It 
reveals a decreasing velocity dispersion with increasing distance from the
M31 center, varying from $\sim122$~km/s at 60~kpc to $\sim57$~km/s at
120~kpc. For reference, the metal-poor field star velocity dispersion
is $\sim$98~km/s at 60~kpc (measured), and $\sim$44~km/s 
at 120~kpc (extrapolated) \citep{Chapman06}.

\subsection{An M31 Mass Estimate}

Assuming the halo GC system is spherically symmetric, we can estimate
the mass of M31 by solving the Jeans equation \citep{BT87}. Because
we have found evidence for rotation in the M31 halo GC system, the
Jeans equation is separated into a rotating and a non-rotating
component. The total mass is obtained by summing the mass supported by
pressure $M_p$ and the rotationally-supported mass $M_{r}$. The
rotational component is determined via:

\begin{equation}
M_{r} = \frac{R_{max}v^{2}_{max}}{G}
\label{eq:rot}
\end{equation}
where $R_{max}$ is the projected radius of the outermost GC in our
sample, $v_{max}$ is the rotational amplitude of the outer GC population
and $G$ is the gravitational constant.

To determine $M_{p}$ we use the solution of the non-rotating Jeans 
equation proposed by \citet{Evans03}, known as the tracer mass 
estimator (TME):

\begin{equation}
M_{p} = \frac{C}{GN}\sum\limits_{i=1}^{N}\left(v_i-v_{sys}\right)^{2}R_i
\label{eq:tme}
\end{equation}
where $R_i$ is the projected radius from the center of M31 for a given
cluster, $v_i$ is the radial velocity of the GC with the rotational
component removed and $N$ is the total number of clusters in our
sample. The constant $C$ depends on the shape of the potential, the
radial distribution of the tracer objects and the anisotropy in the
system. For a spherical, isotropic system, it has the following form:

\begin{equation}
C = {4(\alpha\!+\!\gamma) \over \pi}
    {4\!-\!\alpha\!-\!\gamma\over 3\!-\!\gamma}
    {1\!-\!(r_{\rm in}/r_{\rm out})^{3-\gamma} \over 1\!-\!(r_{\rm in}/r_{\rm out})^{4\!-\!\alpha\!-\!\gamma}}.
\label{eq:C} 
\end{equation}
 
In equation (\ref{eq:C}), $r_{\rm in}$ and $r_{\rm out}$ correspond to 
the smallest and largest 3D radii of the halo GCs respectively. 
In this case, we assume $r_{\rm in} = 30$ kpc,
while $r_{\rm out}$ is set to 200 kpc assuming MGC1 is the most
remote GC \citep{Mackey10MNRAS}. The constant $\alpha$ is related to
the underlying gravitational field, which is assumed to be scale-free,
at least between $r_{\rm in}$ and $r_{\rm out}$. For an isothermal
halo potential where the system has a flat rotation curve at large
radii, $\alpha$ is zero. If we assume a NFW dark matter
profile \citep{NFW}, $\alpha \approx 0.55$ \citep{Watkins10}. The
$\gamma$ parameter is the slope of the GC volume density distribution.
We calculate this using the surface density distribution of all 84 GCs
that have projected radii larger than 30 kpc (Huxor et al. 2013, in
prep; Mackey et al. 2013, in prep) and find $\gamma \sim 3.34 $. It
is also important to note that even though the TME uses a GC sample in
a shell around the center of M31, it calculates the total mass
enclosed by the furthest cluster in that sample.

Using the above method, and setting $\alpha$ to zero, we calculate the
total mass of M31 to be $M_{\rm M31}=1.5\pm0.2\times10^{12}
M_{\odot}$, where the pressure component is $M_{\rm p}=1.3\pm0.2
\times10^{12}M_{\odot}$ and the rotation contribution is $M_{\rm
 r}=2\pm1\times10^{11}M_{\odot}$. Assuming $\alpha = 0.55$,
we find the M31 mass to be $M_{\rm M31}=1.2\pm0.2\times10^{12}
M_{\odot}$, where $M_{\rm p}=1.0\pm0.2\times10^{12}M_{\odot}$ 
and $M_{\rm r}=2\pm1\times10^{11}M_{\odot}$. For reference, 
applying the TME in a single step (ignoring rotation), gives 
$M_{\rm M31}=1.8\pm0.2\times10^{12}M_{\odot}$ and 
$M_{\rm M31}=1.3\pm0.2\times10^{12}M_{\odot}$ for $\alpha$ values 
of 0 and 0.55 respectively. The quoted errors
incorporate the statistical uncertainties only, and in reality they
are much larger. In our mass calculations, we assume isotropic orbits,
a steady state for our tracer population and a power-law form for the
potential. This is the simplest approach we can take, although the 
presence of the substructure in the spatial distribution of the outer 
halo GCs suggests that it may not be correct. Nonetheless, studies 
suggest the presence of substructure in the tracer population will 
bias results only at the 20\% level \citep[e.g.,][]{Yencho06,Deason12}. 
To explicitly test the assumption of steady state, we recalculate the M31 
mass excluding GCs which lie along the North-West stream and stream 
D. We find the total mass of M31 decreases by $0.3 \times 10^{12} 
M_{\odot}$ for both values of $\alpha$, with the formal statistical errors 
remaining unchanged.

\vspace{0.5 mm}
\section{Discussion}

Analysis of the radial velocities of M31 outer halo GCs strongly supports
our earlier finding that many of these objects have been accreted \citep{Mackey10ApJL}. 
Clear kinematic correlations are seen amongst subgroups of GCs which lie on top of stellar
debris features, and, in the case of the North-West stream, an unambiguous signature of radial
infall is observed. Interestingly, recent work has also began to detect 
phase-space substructure in large samples of GCs around distant galaxies \citep[e.g.][]{Strader11,Blom12,
Romanowsky12}. However it is only within the Local Group that we can attempt the obvious 
next step of comparing GC velocities with those of underlying stream stars.
 
A surprising discovery is the high degree of rotation in the M31
outer halo GC population, which is in the same sense as the inner halo
population as well as the main stellar disk. While it is
common to find GCs rapidly rotating in inner regions of galaxies, strong rotation
beyond a few tens of kpc seems to be a rare occurrence, at least amongst early type galaxies 
\citep[e.g.][]{Woodley10,Strader11,Blom12,Pota13}. It is natural to speculate on 
how such coherent motion could arise if the outer GC population were largely
accreted from numerous dwarf galaxy hosts. One possibility 
is that the donor dwarf galaxies have been accreted from a few
preferred directions on the sky and hence have aligned angular momenta, 
as seen in some recent cosmological simulations
\citep{Libeskind05,Libeskind11,Lovell11}. Such a scenario has also been suggested as the origin 
of the planar alignments of dwarf galaxies seen in both the Milky
Way and M31 \citep[e.g,][]{Metz07,ibata13}. Curiously, the plane of dwarf
galaxies reported in M31 rotates in the same sense as the outer halo GC population, although 
the rotation axis of that plane appears to be inclined by $\sim 45\deg$ to the minor axis. 
It is also possible that the bulk of the M31 outer halo GC population was accreted in
a single event involving a moderate mass satellite. Support for this idea could come from
the M31 thick disk which rotates in the same sense, albeit
somewhat faster, than the halo GC population \citep{collins11}. However, 
a rather massive satellite would be required to bring in a population of several tens of 
GCs, raising the question of how the M31 disk could survive such an encounter. 
Furthermore, it would seem difficult to explain the spatial correlation between 
outer halo GCs and the numerous tidal streams in this case. In a different scenario, 
numerical modelling \citep{Bekki10M31} suggests that a past major merger between 
M31 and another disk galaxy could give rise to the rapid rotation of the resulting 
GC system, including the rotation in the halo population.

Despite being our closest massive galaxy, it is remarkable that we are still 
unable to measure the total mass of M31 with good precision. Indeed, there is 
still debate as to whether M31 or the Milky Way is more massive \citep{Watkins10}. 
\citet{Evans03} used GC kinematics to find a M31 mass of $1.2\times10^{12}
M_{\odot}$ out to a deprojected 3D distance of $\sim$100 kpc, while \citet{Galleti06} 
and \citet{Lee08} used expanded samples to calculate masses of $1.9-2.4
\times10^{12}M_{\odot}$ within the same radial range. Several authors 
have attempted to determine the M31 mass via the motions of its dwarf satellite
galaxies \citep[e.g.][]{Cote00,Evans03}. The most recent such measurement is 
presented by \citet{Watkins10} who estimated $1.4\pm0.4\times10^{12} 
M_{\odot}$ using 23 satellites out to 300 kpc, assuming isotropy. Our estimate 
of $M_{\rm M31}=1.2-1.5\pm0.2\times10^{12}M_{\odot}$ within a 3D radius 
of 200 kpc agrees well with this value but suffers from the similar 
systematic uncertainties due to model assumptions. Notably, \citet{vdMarel12} 
use the velocity vector of M31 with respect to the Milky Way in combination with 
the timing argument to derive a total mass for the Local Group $M_{LG} = (4.39
\pm1.63)\times10^{12} M_{\odot}$, which would push the M31 mass higher than 
any estimate thus far using dynamical tracers.

\acknowledgements 
ADM is supported by ARC grant DP1093431.
APH was partially supported by Sonderforschungsbereich SFB
881 ``The Milky Way System" of the German Research Foundation.
The WHT is operated on the island of La Palma 
by the Isaac Newton Group in the Spanish 
Observatorio del Roque de los Muchachos 
of the Instituto de Astrof\'{i}sica de Canarias.

{\it Facilities:} \facility{KPNO (RC-spectrograph) \\} \facility{WHT (ISIS-spectrograph)}.

\end{document}